\documentclass[aps,prl,reprint,groupedaddress]{revtex4-1} 

\usepackage{amsmath,amssymb}
\usepackage{amsfonts}
\usepackage[squaren,cdot]{SIunits}
\usepackage{graphicx}
\usepackage{xcolor}
\usepackage{epstopdf}

\begin{document}

\title{Dynamical coupling between connected foam films by interface transfer across the menisci}
\author{Adrien Bussonni\`ere}
\author{Evgenia Shabalina}
\author{Xavier Ah-Thon}
\author{Micka\"el Le Fur}
\author{Isabelle Cantat}

\affiliation{Univ Rennes, CNRS, IPR (Institut de Physique de Rennes) - UMR 6251, F- 35000 Rennes.}
\date{\today}

\begin{abstract}
The highly confined flow of the liquid phase, trapped between the gas bubbles, is at the origin of the large effective viscosity of the liquid foams. Despite the industrial relevance of this complex fluid, the foam viscosity remains difficult to predict, because of the lack of flow characterization at the bubble scale. Using an original deformable frame, we provide the first experimental evidence of the interface transfer between a compressed film (resp. a stretched film) and its first neighbour, across their common meniscus.
We measure this transfer velocity, which is a key boundary condition for local flows in foams. We also show the dramatic film thickness variation induced by this interface transfer, which may play an important role in the film thickness distribution of a 3D foam sample. 
\end{abstract}


\maketitle

Dry liquid foams are made of thin liquid films connected with each other by menisci. Considering the foam at the millimetric scale of few bubbles, most of the mechanical properties can be understood by modelling the thin films and the menisci respectively as 2D surfaces and 1D lines. The obtained geometrical shape, simply called the foam shape hereafter, is a minimal surface and obeys the Plateau rules \cite{plateau} if the gravity, inertia and viscous forces are negligible, and if the surface tension is homogeneous \cite{livre_mousse_en}.  Under slow imposed deformation, the foam shape follows the evolution of the external constraint, still obeying the Plateau rules: this is the well known elasto-plastic regime \cite{hohler05}.  In this regime, the normal motion of the different films fully determines the mechanical response of the foam; the tangential velocities on the interfaces remain unknown and irrelevant, entirely decoupled from the observed dynamics.  

 At higher deformation rate, in contrast, tangential velocities become crucial, as they are strongly coupled to the viscous forces and surface tension variations which govern the dynamics. In this viscous regime, foam dynamics at the local scale has been partially characterized in the literature:  interfacial stress have been determined by a measure of the angles between adjacent films during imposed deformation \cite{durand06,besson07,biance09,zaccagnino18} or by force measurements \cite{costa13b}; film thicknesses have been measured by interferometry \cite{seiwert13} or light absorption \cite{petit15}; and velocities in a stretched film by particle tracking \cite{petit15}. These measurements provide important information, but are not exhaustive enough to fully determine the flow in the liquid phase. Especially, the amount of interface area which is transferred from one film to its first neighbour during a film deformation, or which is created/destroyed in the meniscus, has never been measured, and is a key parameter in all the current predictions of foam dissipation \cite{besson08, denkov09, costa13,titta18, zaccagnino18}. The principal aim of the paper is to provide the first measurement of the tangential velocities on both sides of a given meniscus, and especially the tangential velocity induced in a film by the deformation of its first neighbour. This measure is synchronized with thickness film and foam shape measurements, during a well controlled deformation. 

\begin{figure}[htp!]
			\centering
   		\includegraphics[width=0.8\linewidth]{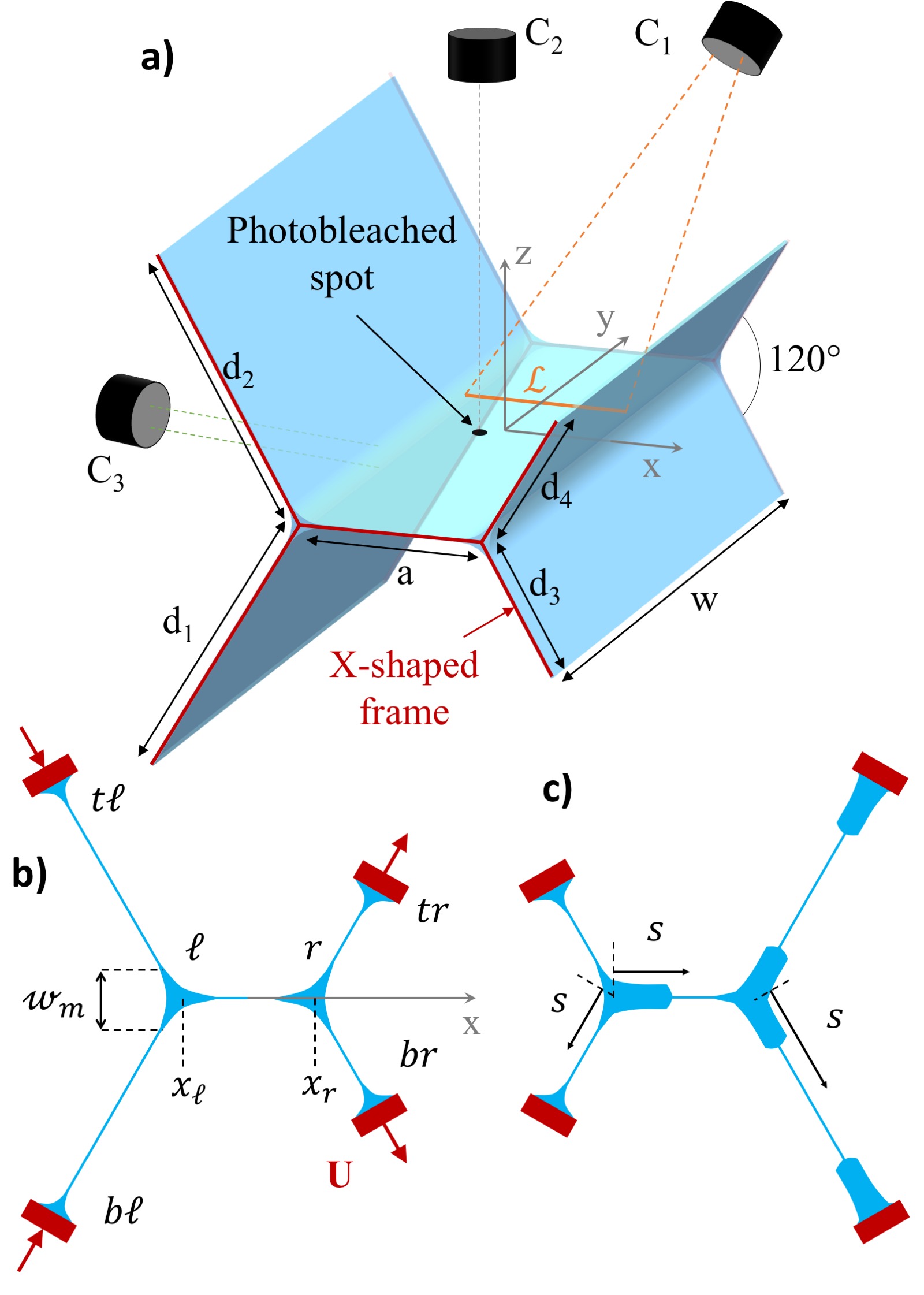}
			\caption{(a) Scheme of the set-up. The 5 foam films and the two free menisci are represented (in blue), as well as the front part of the solid frame (X-shape piece, in red). (b) Representation of the system in the (x,z) plane, at the time $t=0$. The 4 moving edges (top and bottom left $t\ell$, $b\ell$, and top and bottom right $tr$, $br$) are shown in red. (c) Same system, after motor motion. The thick parts of the central and right films, called the Frankel's films, are represented. 
			\label{fig:montage}}
		\end{figure}

A dedicated deformable frame, schematized Fig. \ref{fig:montage}, has been designed on which five connected foam films are produced. The central one has a horizontal rectangular shape; its short edges, of length $a= 6$ mm and oriented along $x$, are the central part of the two X-shaped metallic pieces constituting the lateral boundaries of the system; its long edges, oriented along $y$ are two free menisci of length $w=41.5$ mm, each of them being connected to two other rectangular films, called the lateral films and labelled by the index $i \in [1 ; 4]$ (see Fig. \ref{fig:montage}).

The lateral films make an angle of 120$^o$ with the central film and their external edges are a metallic segment of length $w$ which can translate along the lateral arms of the X-shaped pieces. The distance $d_i(t)$ between the moving edge $i$ and the corresponding meniscus is controlled by a linear piezo motor (PI U-521.23). This geometry has been chosen so that each film remains flat and stay in the same plane during the deformation.  In this study, we impose $d_{1,2}=d_\ell= d - U t$ on the left side and $d_{3,4} = d_r= d+ U (t -\tau)$  on the right side, between $t=0$ and $t=\tau$. The four motors stop after a displacement of $10$ mm, at time $\tau = 0.2$ s. The motor velocity is $U=50$ mm/s and $d=7.1$ mm. Syringes positioned along the four upper arms of the X-shaped pieces each injects a flux $Q = 0.05$ mL/min of surfactant solution in the system. This allows to prepare a reproducible initial state for the films and menisci, which are deformed 15 s after being pulled at a controlled velocity out of the foaming solution. 

The  solution is made of deionised water, sodium dodecyl sulfate (SDS, cmc = $2.8$ g/L) at $5.6$ g/L, dodecanol at $0.05$ g/L, glycerol at $15\ \%$ in volume and fluorescein at $0.8$ g/L. The viscosity is $\eta = 1.52$  mPa.s, and the equilibrium surface tension is $\gamma= 32$ mN/m.  

The most extensive set of data has been obtained in the central film, and reveals that the deformation of the lateral films induce an important flow in the central film, as well as a strong swelling, even if the central film area is not modified. This unexpected dynamical behavior has been fully characterized by 3 synchronized cameras.  

The central film thickness $h(x,t)$ is measured along a line $\cal{L}$ oriented along $x$, with a spectral camera Resonon Pika L (camera $C_1$) used at a frame rate of $200$ im/s. This camera records light spectrum $I(\lambda)$ along the vertical lines of the sensor, each point of $\cal{L}$ corresponding to a different sensor line (see Supp. Mat. A \cite{SM}). The thickness $h(x,t)$ is extracted by using the relation $I(\lambda) \propto 1-\cos(4\pi h(n^2-\sin^2{\theta})^{1/2}/\lambda)$, with $n=1.33$ the water refractive index and $\theta=45^o$ the incidence angle.

Film thickness profiles have also been obtained in the bottom lateral films, hereafter denoted by the left and right films. 
 
The velocity field is deduced from the motion of photobleached spots \cite{seiwert17}. A laser of wavelength $488$ nm and power $200$ mW is focused on the film with a normal incidence, during $15$ ms. 
The fluorescein is photobleached over the whole film thickness in a cylindrical region of diameter $200\ \mu$m.  
A large part of the film is illuminated at the same wave length and the light of wavelength between $505$ and $545$ nm emitted by the fluorescein is recorded with a camera placed above the setup  (camera $C_2$). For each experiment, a single spot is photobleached 10 ms before motor motion, and its position is followed during few seconds (unless it reaches the central film boundaries before).
The velocity remains uniform across the central film thickness because (i) the pressure is the ambient pressure in the flat part of the film,  preventing any Poiseuille flow; (ii) the symmetry of the system and of the imposed deformations implies identical velocities on both interfaces, preventing any shear flow. Moreover the experimental time is shorter than the fluorescein diffusion time. Consistently the spot remains well contrasted along its whole trajectory and is a good passive tracer. 

This camera $C_2$ is also used to track the positions  $x_\ell(t)$ and  $x_r(t)$, of the left and right menisci, respectively. They slightly move during the deformation, without modifying significantly the central film area.  In the central film, we denote by $s$ the distance  to the left meniscus, {\it i.e.} $s = x- x_\ell(t)$, while in the lateral films $s$ is the distance to the corresponding free meniscus (see Fig. \ref{fig:montage})c.

Finally, the vertical dimension $w_m$ of the left meniscus is recorded by light transmission using a collimated white light and a telecentric objective Edmund optic Silver-TL 4x (camera $C_3$). The vertical position of the meniscus remains almost constant, with variations less than $100\ \mu$m.

 \begin{figure}[htp!]
		\begin{center}
	\includegraphics[width=0.85\linewidth]{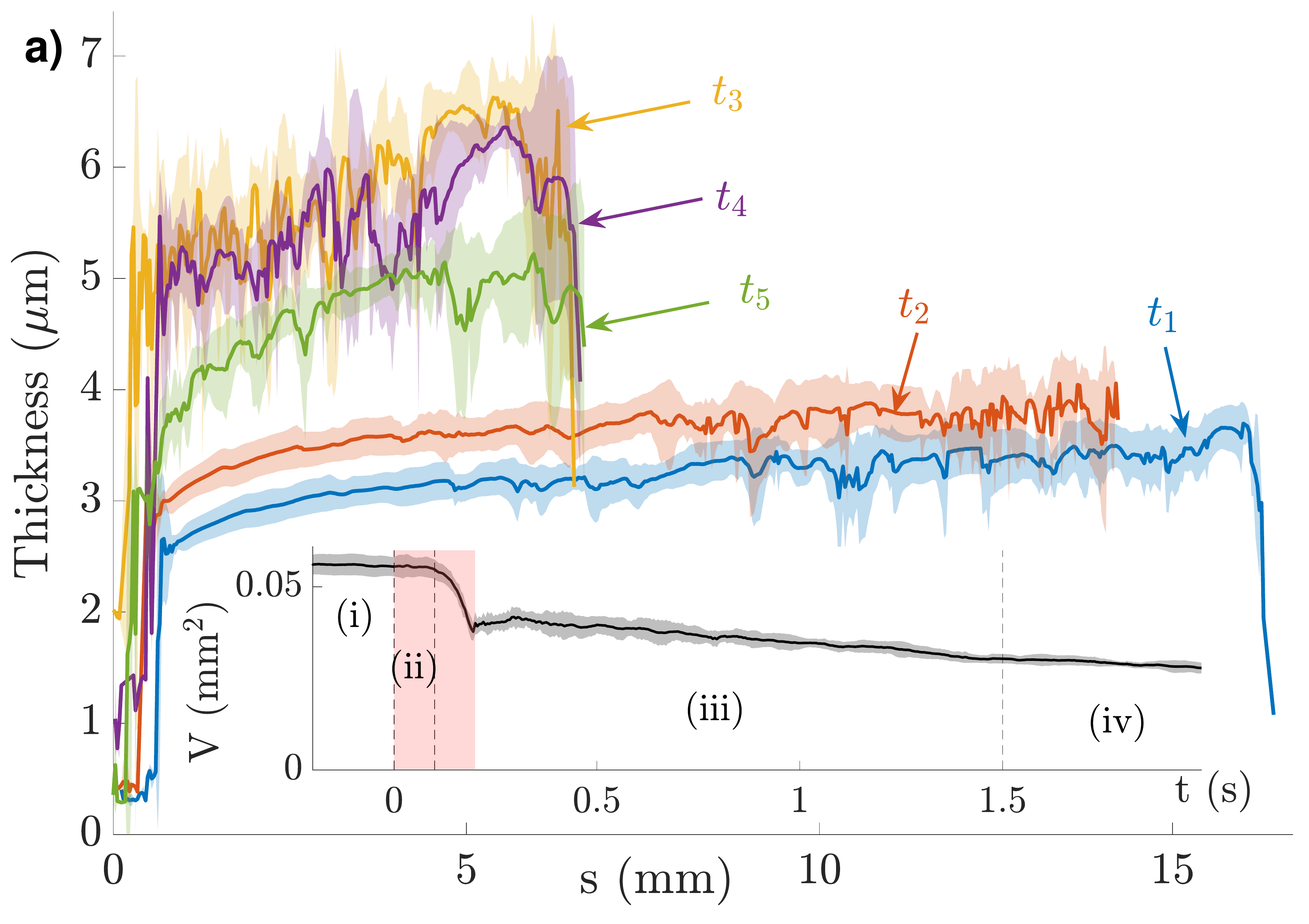}\\
 	\includegraphics[width=0.85\linewidth]{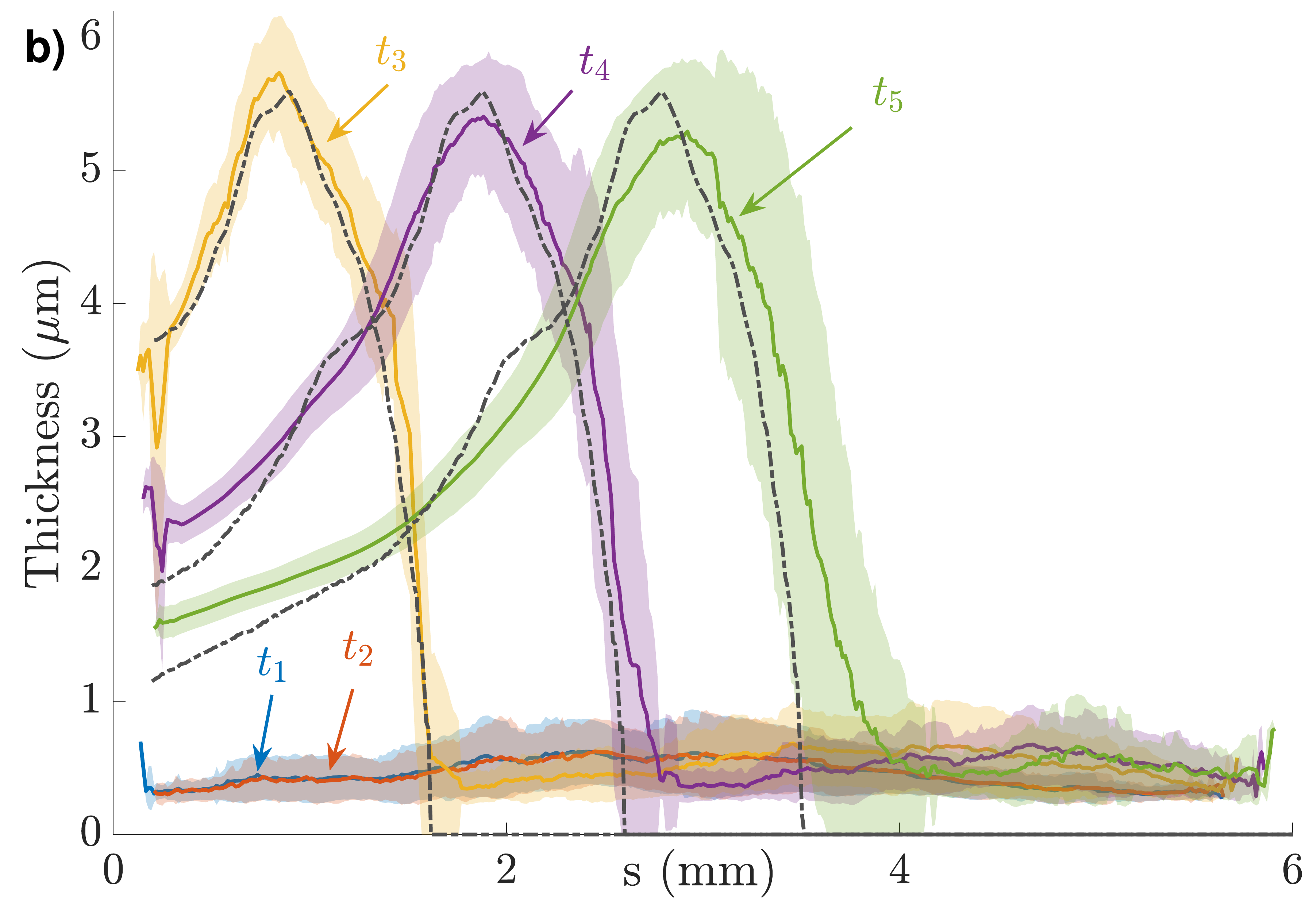}\\
	\includegraphics[width=0.85\linewidth]{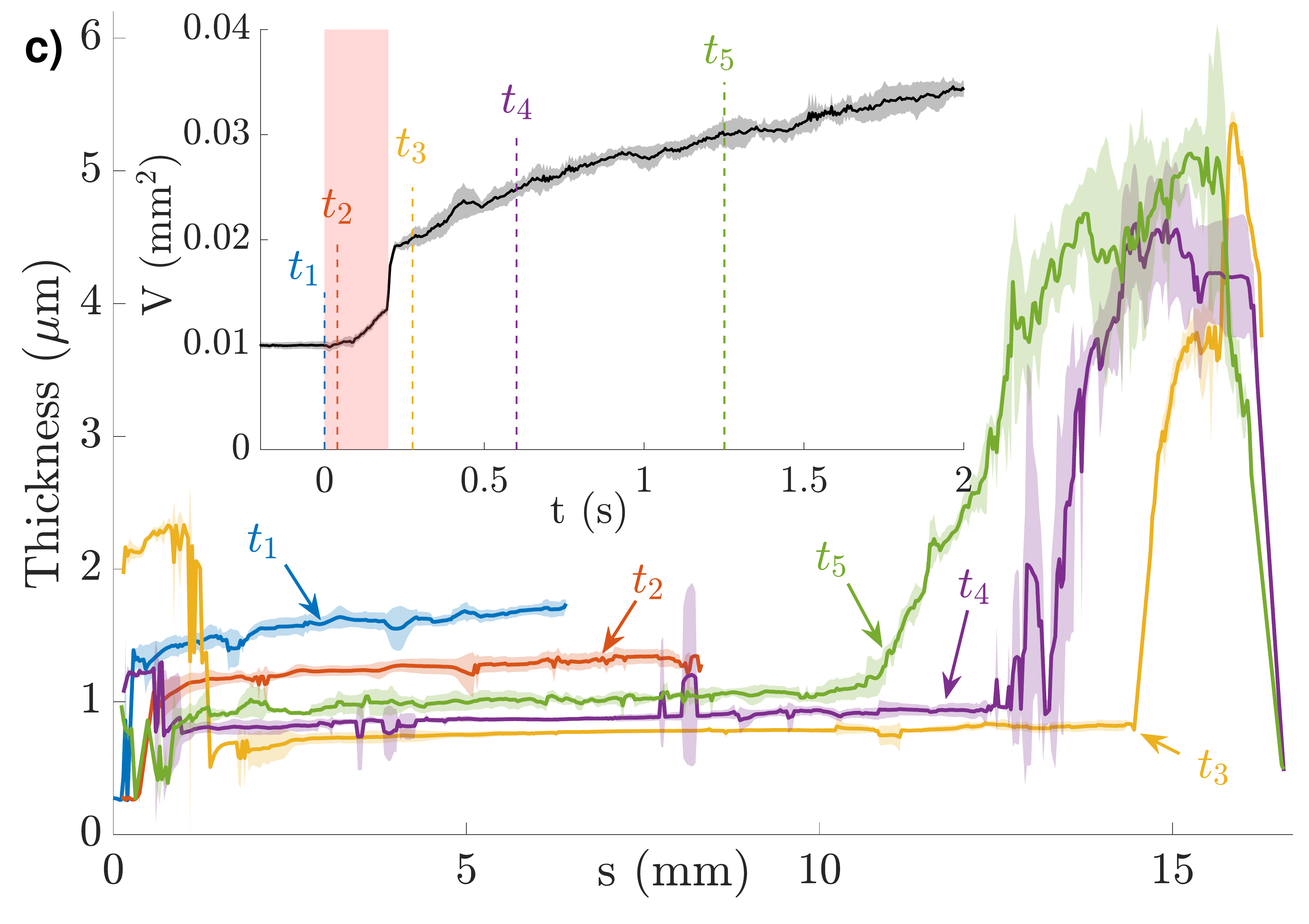}
			\caption{Film thickness profiles in the films at $t_1= 0$ s (step  i); $t_2=0.04$ s (step ii);  $t_3 = 0.075$ s (step iii, before the gravitational instability) and   $t_{4,5}= [0.6, 1.25]$ s (step iii, after instability). The solid lines (resp. shaded areas) represent the average (resp. standard deviation) of 5 experiments for the lateral films and 50 for the central.
		 (a) Bottom left film (compressed, upper meniscus at $s=0$); (b) Central film (left  meniscus at $s=0$). The dashed lines correspond to the thickness profiles predicted by equation \ref{frankel} ; (c) Bottom right film (stretched, upper meniscus at $s=0$).
		 The insets show the left (a) and right (c) film volumes (per unit length in the $y$ direction) $V_\ell$ and $V_r$, and indicates  the times $t_i$, times domains (i - iv) and  motor motion time (shaded red).
			\label{fig:profiles}}
			\end{center}
		\end{figure}

The  thickness profiles  during and after motor motion  are shown in Fig. \ref{fig:profiles} for different times, in the central, left and right films, evidencing the following phenomenology :

(i) {\it Initial state} $(t<0)$. The central film thickness is uniform and equal to $h_0 = 400 \pm 200$ nm. The left film is 3 $\mu$m thick and the right film is 1.5 $\mu$m thick. These last thicknesses have steady but out of equilibrium values, which depend on the film size and inclination, and on the injected flux. The meniscus radius of curvature is deduce from the equilibrium relationship $r_m= w_m = 320\ \mu$m \cite{livre_mousse_en}.

(ii) {\it Lateral films stretching and compression} ($t\in [0 - 0.1]$ s). The right film get thinner and the left film get thicker, but their volumes remains constant. This evidences a homogeneous films deformation with no exchange with the menisci. No noticeable evolution is observed in the central film.

(iii) {\it Interface transfer across menisci} ($t\in [0.1 - 1.5]$ s).
In the right film, thick pieces of film, called the Frankel's films, are extracted from both the right and bottom right menisci (see Fig. \ref{fig:montage}c).  A straight, well defined, frontier separates these Frankel's films from the film initially present (hereafter called the  initial film). The total film volume increases, while the initial film keeps a constant volume, and a homogeneous and increasing thickness; symmetrically, the left film get thinner and its volume decreases, evidencing a liquid flux toward the left and bottom left menisci. A similar behavior is observed in the top lateral films.
Note that during steps (ii) and (iii) the thickness in the left film and in the right initial  film evolves non-monotonically, as shown in Fig. 1 of the Sup. Mat. \cite{SM}.
In the central film, a Frankel's film is extracted from the left meniscus and the thin film initially present flows into the right meniscus. This step (iii) begins while the motors are still moving, but lasts for few second after motors stop at $t=0.2$ s. 
The top/bottom symmetry of the motion in the right film is broken at $t \approx 0.25$ s by a gravitational instability which imposes a stratification of the film thickness, thicker parts of the film laying below the thinner parts. A single Frankel's film is thus observed at later times in the right film, close to the bottom meniscus. The Frankel's film is however continuously produced at the top, but it falls down after its extraction from the meniscus and merges with the Frankel's film produced at the bottom (see \cite{shabalina18}). The downwards motion of patches of thick films induces recirculations in the central part of the stretched film,  thus forbidding reliable velocity measurements in this lateral film. 
  
(iv) Eventually, the system recovers its slow drainage behavior, after a delay of the order of 10 s.

The Frankel's film extraction in the central film is due to the motion of the interfaces across the left and right  menisci, which we determine  from the velocity measurements. We measured the spots position relatively to the left meniscus  $s(t) = x(t)- x_\ell(t)$. The displacement of the spots in the $y$ direction remains smaller than $10\ \%$ of the total spot displacement, except for positions very close from the right meniscus, where we observe marginal regeneration which locally breaks the invariance in the $y$ direction \cite{mysels}. In the following, the spots at a distance smaller than 1 mm from the right meniscus are discarded. We recorded 4 trajectories $s_i(t)$, each being averaged over 10 experiments. The index $i\in[1-4]$ refers to the initial distance $s_i(0)$ between the spots (i) and the left meniscus, ranging from 1 to 4 mm.

\begin{figure}[htp!]
			\centering
   		\includegraphics[width=0.85\linewidth]{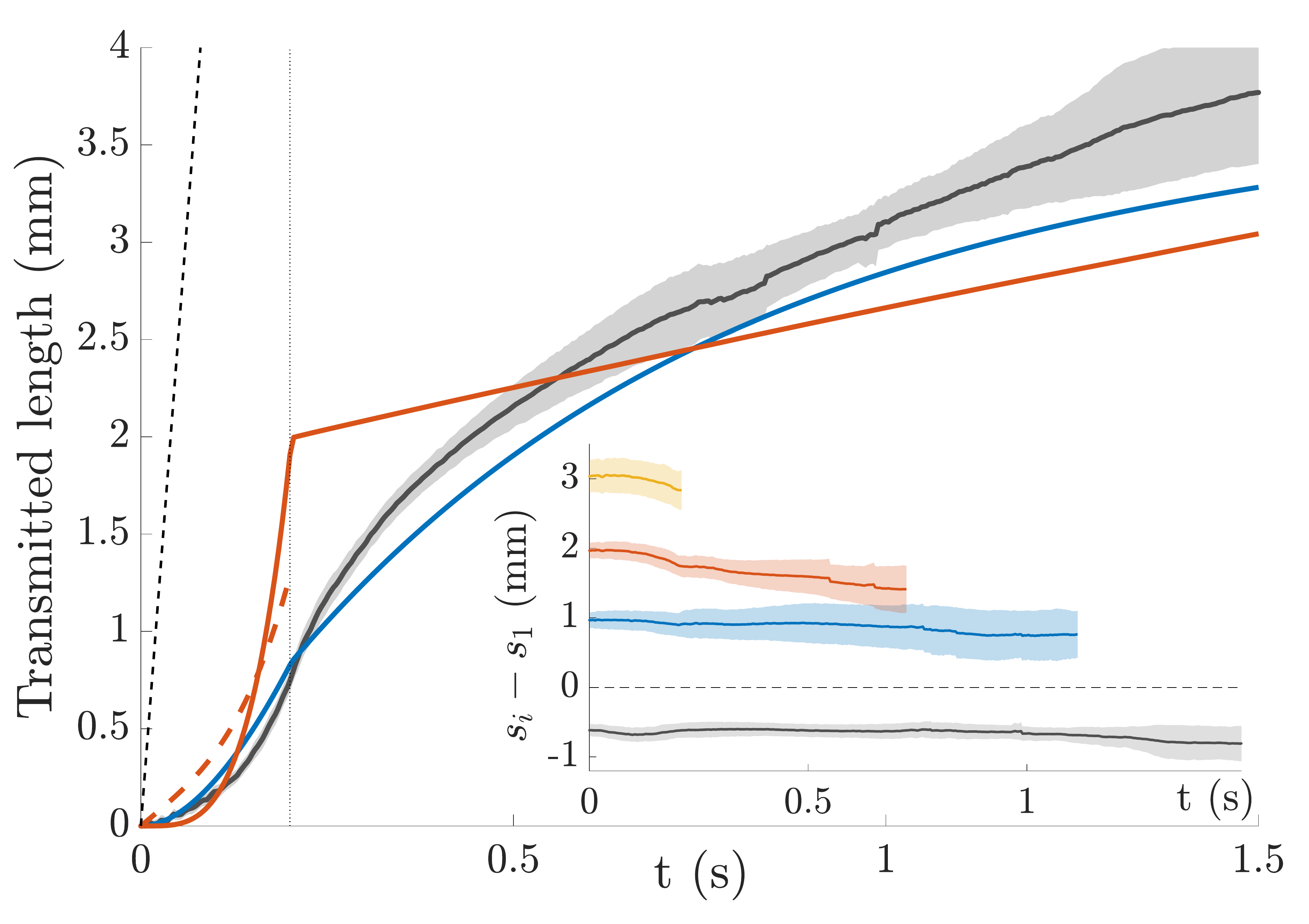}
			\caption{Amount of interface $L$ transferred from the compressed side to the stretched side as a function of time. Black, full line: $L_c(t)$ (central film) ;  Red, full line: $L^\ell$; dashed line:  $L^\ell_{sym}$ (compressed film) ;  Blue:  $L^r$ (stretched film). The dashed black line shows the motor displacement, stopping at the dotted black line. Inset: trajectories of the spots (2-4, colored) and of the frontier of the Frankel's film (black) relatively to the spots (1). 
				 Errors bars are the standard deviation for 10 trajectories.
			\label{fig:spots}}
		\end{figure}

These trajectories evidence a translational motion of the central film, with almost no deformation, over a distance which is a significant fraction of the motor displacement. Fig. \ref{fig:spots}(inset) shows the distance between spots $[(2) - (4)]$ and the spots $(1)$ as a function of time. The spots $(1)$ and $(2)$ are close to the left meniscus, and the distance between them is almost conserved during the whole motion, showing the absence of film deformation in this region. 
The distances between spots (3,4) and spots (1) slightly decrease, indicating a small film compression in the right part of the film. However, no clear signature of this compression is observable in the thickness map (Fig. \ref{fig:profiles}b), and residual recirculations close to the right meniscus may be at the origin of this small drift, especially at the longest times.

The trajectory of spots $(1)$ relatively to their initial position, {\it i.e.}  the  distance $L^c(t) = s_1(t) -s_1(0)$, is plotted in Fig. \ref{fig:spots}. As the central film moves as a whole close to the left meniscus,  $L^c$ corresponds to the amount of film extracted from the left meniscus.  If we neglect the small compression observed close to the right meniscus, it corresponds also to the amount of film absorbed by the right meniscus. 

In the left and right films, the interface length transmitted across the meniscus $L^{r,\ell}(t)$ is determined from the thickness profile. On both sides, we assume  that the same amount of interface goes in (or out) the top and bottom meniscus bounding the lateral film. In that case,  we establish in Supp. Mat. B \cite{SM} that:
\begin{equation}
v^{sym}_{r,\ell}(0,t) =\frac{1}{2} \left( \dot{d}_{r,\ell} - d_{r,\ell}\frac{\dot{\varepsilon_{r,\ell}}}{\varepsilon_{r,\ell}}  \right )\;  ,
\label{v0sym}
\end{equation}
  with $v^{sym}_{r,\ell}(0,t)$ the film velocity at the free meniscus and $\varepsilon_{r,\ell} = \langle h_{r,\ell} \rangle_{(t=0)}/\langle h_{r,\ell}\rangle_{(t)}$ the film extension. The average $\langle \rangle$ is made respectively on the whole left film, and on the thin part of the right film. Then we define  $L^{r,\ell}_{sym}(t)= - \int_0^t v^{sym}_{r,\ell}(0,t) dt$.

 In the right film, the top and bottom Frankel films have the same width before the instability begins, thus confirming the validity of our symmetry assumption, at least  at short time.
In the left film, a spot trajectory $s_\alpha(t)$ has been obtained, again at short time, allowing to compute $v_\ell$ as  $v_\ell(0,t)=\dot{s}_\alpha-\frac{\dot{\varepsilon}}{\varepsilon} \, {s_\alpha}$, which does not requires any symmetry assumption (see Supp. Mat. B \cite{SM}). The prediction  $L^\ell(t)= - \int_0^t v_\ell(0,t) dt$ is in qualitative agreement with $L^\ell_{sym}(t)$, as shown in Fig. \ref{fig:spots}.

The quantity $L^c$ is determined with an error bar smaller than 10 $\%$. $L^r$ and $L^\ell$ (or $L^\ell_{sym}$) have a larger uncertainty at large time, due to the assumption of symmetric behavior we need to make. However, they show a behavior very similar to $L^c$ with time (Fig. \ref{fig:spots}). A first conclusion is thus that the motor displacements lead to the transfer of interface from the left film to the right film across the two free menisci and the central film, occurring with negligible in-plane deformation, and with some delay with respect to the motor motion. The amount of interface transferred after 1 s is of the order of 3 mm, which is a significant part of the 10 mm displacement of the motor.

This interface transfer across the central film is coupled to the extraction of a Frankel's film from the left meniscus. After few seconds, it entirely invades the central film which is, thus, strongly modified by the deformation, even if its area remains unchanged: 
the extracted film is one order of magnitude thicker than the film initially present. It is separated from the thin film by a sharp frontier, moving at the velocity $v_c(t)= dL^c/dt$ (see Fig. \ref{fig:spots} inset). 

The thickness of the film extracted from a meniscus of radius $r_m$, at a constant velocity $v$ is given by the Frankel's expression $h^{Fr} = 2.66\,  r_m Ca^{2/3}$, with $Ca= \eta v/ \gamma$ \cite{mysels}. In the central film, the dynamical meniscus lateral extension scales as $\delta_{dyn}= r_m Ca_c^{1/3} \sim 10\ \mu$m and the film goes through this region in a time $t_{dyn} \sim \delta_{dyn}/v_c\sim 10^{-3}$ s, much smaller than the time scale of the  velocity variation $v_c /(dv_c/dt) \sim 0.1$ s (see Fig. \ref{fig:spots}). An approximation of quasi steady motion is thus justified.  
 The thickness gradient at the frontier is of the order of $\phi\approx 10^{-2}$ and it decreases on the typical time scale $\phi^4 \eta h /\gamma \approx 10$ s \cite{chai14}, the diffusion of the step-like frontier thus occurs on a slower timescale than the experiment.
 Once extracted from the meniscus, the film thus simply follows a passive convection.
Moreover, as the central film extension rate is negligible, it moves without any deformation.
As a consequence, the film  at the position $x$ at time $t$ has been extracted from the meniscus at time $t_0(x)$ implicitly given by the relation  $x = \int_0^{t_0(x)} v_c(t) dt $, and the theoretical prediction for its thickness is 
\begin{equation}
h^{th}(x,t)= 2.66 \,  r_m \left ( \frac{\eta v_c(t_0(x))}{\gamma} \right)^{2/3} \; .  
\label{frankel}
\end{equation}
The quantitative agreement shown in Fig. \ref{fig:profiles} between this prediction, obtained without adjustable parameter, and the experimental data, validates {\it a posteriori} the assumptions of the Frankel's model.

In conclusion, the imposed deformation induces a thickness increase by a factor of 10 in a film which is only a neighbour of the deformed films. The thickness relaxation toward its initial value last for several seconds. We believe that the localized deformations occurring in 3D foam samples, due to aging or global flow, may redistribute the liquid phase in the thin films at a much larger distance from the local deformation than previously expected. This process is potentially an efficient factor of rejuvenation of the foam films, competing with the slow drainage imposed by the meniscus capillary suction and gravity.  
The original set-up we developed proved, on the example of the imposed deformation we choose, its capacity to identify deformation modes in foam films, and to measure them with an accuracy allowing for a quantitative comparison with local theoretical models. We believe this should remains the case for a large class of imposed deformations, which paves the way to a deeper understanding of dissipation in foams.    

\begin{acknowledgments}
This project has received funding from the European Research Council (ERC) under the European Union’s Horizon 2020 research and innovation program (grant agreement No 725094). 
 We thank A. Saint-Jalmes and A. B\'erut for fruitful discussions and E. Schaub for technical support. 
\end{acknowledgments}

\bibliography{bib}
\bibliographystyle{apsrev4-1} 

\end{document}